\documentclass{article}
\usepackage{PRIMEarxiv}

\usepackage[utf8]{inputenc}     
\usepackage[T1]{fontenc}        

\usepackage{hyperref}           
\usepackage{url}                
\usepackage{booktabs}           
\usepackage{amsmath, amsfonts}  
\usepackage{microtype}          
\usepackage{fancyhdr}           
\usepackage{graphicx}           

\usepackage{adjustbox}
\usepackage{tikz}
\usetikzlibrary{quantikz2}

\usepackage[
  backend=biber,
  style=numeric,
  sorting=none]{biblatex}   
\title{Generalised Quantum Gates for Qudits and their Application in Quantum Fourier Transform}

\author{
  Francesco Pudda \\
  \texttt{francesco.pudda@mail.com} \\
  \AND
  Mario Chizzini \\
  Università di Parma \\
  \texttt{mario.chizzini@unipr.it} \\
  \And
  Luca Crippa \\
  \texttt{luca.crippa88@gmail.com} \\
}

\begin{document}
\maketitle

\begin{abstract}
  Quantum computing with qudits, quantum systems with $d > 2$ levels, offers a powerful extension beyond qubits, expanding the computational possibilities of quantum systems, allowing the simplification of the implementation of several algorithms and, possibly, providing a foundation for optimised error correction. In this work, we propose a novel formulation of qudit gates that is universally applicable for any number of levels $d$, without restrictions on the dimensionality. By extending the mathematical framework of quantum gates to arbitrary dimensions, we derive explicit gate operations that form a universal set for quantum computation on qudits of any size. We demonstrate the validity of our approach through the implementation of the Quantum Fourier Transform (QFT) for arbitrary $d$, verifying both the correctness and utility of our generalized gates. This novel methodology broadens the design space for quantum algorithms and fault-tolerant architectures, paving the way for advancements in qudit-based quantum computing.
\end{abstract}


\section{Introduction}
Quantum computing has emerged as a transformative technology with the potential to solve certain computational problems with an exponential speed-up with respect to classical computers \cite{nielsen2001quantum}. While most research has focused on qubits, the fundamental two-level units of quantum information, there has been growing interest in exploiting quantum systems with higher dimensions, known as qudits \cite{gottesman2001encoding, thew2002qudit, sheridan2010security, li2013geometry, chi2022programmable, ringbauer2022universal}.

A qudit, standing for \textit{quantum digit}, represents a quantum state with $d$ levels where $d > 2$, generalizing the concept of a qubit (which is limited to two levels). Unlike qubits, which can exist in a superposition of $0$ and $1$, qudits can exist in a superposition of multiple states, such as $0,\, 1,\, \ldots,\, d-1$. They can be physically realized using systems with naturally higher energy levels, such as trapped ions \cite{low2020practical, ringbauer2022universal}, photonic systems \cite{chi2022programmable, karacsony2024efficient}, molecular nanomagnets \cite{chizzini2022quantum, chizzini2022molecular, chiesa2024molecular}, or by accessing higher levels on superconducting transmons \cite{liu2023performing, fischer2023universal, wang2024systematic}. This expanded computational capacity enables the execution of multiple control operations in parallel, thereby potentially reducing circuit complexity, streamlining experimental setups, improving algorithmic efficiency, and, possibly, embedding quantum error correction algorithms within the qudits themselves \cite{li2013geometry, amin2013adiabatic, chizzini2022quantum}.

The advantages of qudit-based quantum computing are manifold. By encoding information in higher-dimensional state spaces, qudits can potentially require fewer quantum resources to perform certain computations, leading to more efficient quantum circuits \cite{li2013geometry, cruz2021towards}. This efficiency comes from the ability of qudits to store more information per quantum unit, reducing the number of gates and operations required in complex algorithms. Additionally, qudits have been shown to offer improved resilience against certain types of noise and errors, a critical challenge in realizing fault-tolerant quantum computers \cite{li2013geometry, amin2013adiabatic, chizzini2022quantum, chizzini2022molecular}. This resilience stems from their ability to exploit a larger set of quantum states, offering more flexibility in error correction strategies and enhancing robustness in noisy environments.

Despite these promising prospects, the development of a comprehensive theory of qudit quantum gates has not been thoroughly explored in a general context. In this work, we introduce a general set of definitions and formulations that extend the standard quantum gate theory to qudits of arbitrary dimensionality, and we prove that a qudit-based Quantum Fourier Transform (\textit{QFT}) can be built by applying these definitions. Our results offer an alternative foundation for designing qudit-based algorithms and architectures, offering new directions for both theoretical research and practical applications in quantum computing.

\section{Quantum qudit gates}
As in classical computation, the concept of a universal set of gates is also crucial in quantum computing, whether dealing with qubits or qudits. A universal gate set ensures that any quantum operation can be decomposed into a sequence of these fundamental gates, much like how classical logic gates can represent any classical computation. Without a universal set, certain quantum transformations would be inaccessible, limiting the computational power of a quantum system. Moreover, having a well-defined universal gate set simplifies the design and optimization of quantum algorithms, enabling more efficient execution on quantum hardware—particularly important given the noise and error rates of current devices. In this section, we will explore the foundational qudit gates, such as Pauli and Hadamard gates, as well as controlled gates, and define some key Z-axis rotation gates.

\subsection{The Pauli and Hadamard gates}
A generalised Clifford algebra in higher dimensions is well-studied in the literature \cite{tesser1992generalized, tao1995generalized, granik1996new}, and its applications to quantum computing provide the cornerstone of quantum mechanical dynamics in finite-dimensional vector spaces \cite{weyl1927quantenmechanik}. From the two-dimensional case, multiple generalised forms are possible \cite{patera1988pauli, gottesman2001encoding}, but here we will focus on Schwinger's canonical notations \cite{ramakrishnan1971generalized}.

The gates from the Pauli group are commonly referred to, as the \textbf{\textit{X}}-gate and \textbf{\textit{Z}}-gate or, respectively, the \textbf{\textit{Shift}} operator and \textbf{\textit{Clock}} operator. From these gates we can derive the formulation of the Hadamard gate, or \textbf{\textit{H}}-gate. For $d$-dimensional qudits, their matrix forms are:

\begin{equation}
  X_d = \begin{bmatrix}
    0 & 0 & \ldots & 0 & 1 \\
    1 & 0 & \ldots & 0 & 0 \\
    0 & 1 & \ldots & 0 & 0 \\
    \vdots & \vdots & \ddots & \vdots & \vdots \\
    0 & 0 & \ldots & 1 & 0 \\
  \end{bmatrix}
  \quad Z_d = \begin{bmatrix}
    1 & 0 & 0 & \ldots & 0 \\
    0 & \omega & 0 & \ldots & 0 \\
    0 & 0 & \omega^2 & \ldots & 0 \\
    \vdots & \vdots & \vdots & \ddots & \vdots \\
    0 & 0 & 0 & \ldots & \omega^{d-1} \\
  \end{bmatrix}
  \quad H_d = \frac{1}{\sqrt{d}} 
  \begin{bmatrix}
    1      & 1            & 1               & \cdots & 1 \\
    1      & \omega^{d-1} & \omega^{2(d-1)} & \cdots & \omega^{(d-1)^2} \\
    1      & \omega^{d-2} & \omega^{2(d-2)} & \cdots & \omega^{(d-1)(d-2)} \\
    \vdots & \vdots       & \vdots          & \ddots & \vdots \\
    1      & \omega       & \omega^2        & \cdots & \omega^{d-1}
  \end{bmatrix}
\end{equation}

Where $\omega$ is the unit root, $\omega = e^{i \frac{2 \pi}{d}}$. The above matrices result in the following actions:

\begin{equation}
  X_d \left | j \right \rangle = \left | \left ( j + 1 \right ) \bmod{d} \right \rangle
  \quad Z_d \left | j \right \rangle = \omega^j \left | j \right \rangle
  \quad H_d \left | j \right \rangle = \frac{1}{\sqrt{d}} \sum_{k=0}^{d-1} \omega^{- j k} \left | k \right \rangle
\end{equation}

The \textbf{\textit{X}}-gate shifts up the qudit's state cyclically by one level, while the \textbf{\textit{Z}}-gate applies a phase rotation dependent on the state level. In particular, it rotates $j$-times the phase of the qudit by $\omega$, clockwise around the \textbf{\textit{Z}}-axis, where $j$ is the current qudit state. On the other hand, the \textbf{\textit{H}}-gate induces a superposition across qudit states, with each state phase-shifted by a multiple of $\omega$. For $d = 2^k$, the $H^{\dagger}$ coincides with the discrete Fourier transform, converting position coordinates to momentum coordinates and vice versa. $H$ is related to $X$ and $Z$, by the following relation:
\begin{equation}
  X = H Z H^{\dagger}
\end{equation}

\subsection{The Z-rotation gates}
Let's now define other common gates, in particular, gates that rotate the phase of the qudit around the \textbf{\textit{Z}}-axis. The first one is the \textbf{\textit{S}}-gate, also called "half" \textbf{\textit{Z}}-gate or $\boldsymbol{\sqrt{Z}}$-gate. The gate is part of the Clifford group and, here, we define it as the operation whose effect is identical to the previously defined \textbf{\textit{Z}}-gate, with half its rotations.

Nevertheless, from the Gottesman-Knill theorem \cite{gottesman1998theory}, it is shown that the Clifford and Pauli gates alone do not guarantee universal quantum computing. We need then to define another gate such that, combined with the previously defined ones, allows for any rotation around the Bloch sphere: the \textbf{\textit{T}}-gate. The \textbf{\textit{T}}-gate is not part of the Clifford group and, similarly to the \textbf{\textit{S}}-gate, can be seen as "half" of it, or a "quarter" of a \textbf{\textit{Z}}-gate. Therefore it is also called the $\boldsymbol{\sqrt{S}}$-gate or the $\boldsymbol{\sqrt[4]{Z}}$-gate. We define their matrix forms as:

\begin{equation}
  S_d \doteq \begin{bmatrix}
    1 & 0 & 0 & \ldots & 0 \\
    0 & \omega^{\frac{1}{2}} & 0 & \ldots & 0 \\
    0 & 0 & \omega & \ldots & 0 \\
    \vdots & \vdots & \vdots & \ddots & \vdots \\
    0 & 0 & 0 & \ldots & \omega^{\frac{d-1}{2}} \\
  \end{bmatrix}
  \quad T_d \doteq \begin{bmatrix}
    1 & 0 & 0 & \ldots & 0 \\
    0 & \omega^{\frac{1}{4}} & 0 & \ldots & 0 \\
    0 & 0 & \omega^{\frac{1}{2}} & \ldots & 0 \\
    \vdots & \vdots & \vdots & \ddots & \vdots \\
    0 & 0 & 0 & \ldots & \omega^{\frac{d-1}{4}} \\
  \end{bmatrix}
\end{equation}

Given the above definitions, we propose a novel but simple form to define both as a function of the rotation. In fact, as the qubit \textbf{\textit{P}}-gate is a generalisation of any rotation around the \textbf{\textit{Z}}-axis, we can extend its form to the qudit case. Its effect is then straightforward: a $\boldsymbol{P(\theta)}$-gate, applied to a qudit in a state $j$, rotates the qudit's phase $j$-times by $\theta$ radians clockwise around the \textbf{\textit{Z}}-axis, similarly to the \textbf{\textit{Z}}-gate. In the matrix form below, we expressed $\theta$ as a function of $\omega$ for a better comparison with the other gates.

\begin{equation}
  P_d(\theta) \doteq \begin{bmatrix}
    1 & 0 & 0 & \ldots & 0 \\
    0 & \omega^{\frac{\theta}{\pi}} & 0 & \ldots & 0 \\
    0 & 0 & \omega^{\frac{2 \cdot \theta}{\pi}} & \ldots & 0 \\
    \vdots & \vdots & \vdots & \ddots & \vdots \\
    0 & 0 & 0 & \ldots & \omega^{\frac{(d - 1) \cdot \theta}{\pi}} \\
  \end{bmatrix}
\end{equation}
\begin{equation}
  P_d(\theta) \left | j \right \rangle \doteq \omega^{\frac{j \cdot \theta}{\pi}} \left | j \right \rangle
\end{equation}

\subsection{The controlled gates}
In order to complete a set of quantum gates, suitable to get to a universal set, we need to create entanglement between qudits.

Let's first consider the \textbf{\textit{CX}}-gate. Differently from the qubits case, where it acts as a do-or-don't gate, its general definition must take into account the fact that there may be multiple "positive" levels. Indeed, in literature, the general form for arbitrary $d$ will not shift up the level of the target qudit by one if a certain condition is met, but will shift it circularly up $j$-times, where $j$ is the state of the controlled qudit \cite{howard2012qudit, garcia2013swap, prakash2021normal}. That's why it is also called the \textbf{\textit{Sum}} operator. Here, we are going to call the gate, the \textbf{\textit{SUMX}}-gate.

\begin{equation}
  SUMX_d = \begin{bmatrix}
    I_d &      &       &        & \\
        & X_d  &       &        & \\
        &      & X_d^2 &        & \\
        &      &       & \ddots & \\
        &      &       &        &  X_d^{d-1} \\
  \end{bmatrix}
\end{equation}
\begin{equation}
  SUMX_d \left | j \right \rangle \left | k \right \rangle = \left | j \right \rangle \left | \left ( j + k \right ) \bmod{d} \right \rangle
\end{equation}

In the same way, we can define a gate for a controlled rotation around the \textbf{\textit{Z}}-axis: the \textbf{\textit{SUMP}}-gate. In fact, by taking into account the previously defined \textbf{\textit{P}}-gate and \textbf{\textit{SUMX}}-gate, we can infer its matrix form as:

\begin{equation}
  SUMP_d \left ( \theta \right ) \doteq \begin{bmatrix}
    I_d &      &       &        & \\
        & P_d \left ( \theta \right )  &       &        & \\
        &      & P_d^2 \left ( \theta \right ) &        & \\
        &      &       & \ddots & \\
        &      &       &        &  P_d^{d-1} \left ( \theta \right ) \\
  \end{bmatrix}
\end{equation}

Its effect is straightforward: it rotates the phase of the target qudit just like the \textbf{\textit{P}}-gate, by an amount proportional to the product of the levels of the control and target qudits.

\begin{equation}
  SUMP_d (\theta) \left | j \right \rangle \left | k \right \rangle \doteq \omega^{\frac{j \cdot k \cdot \theta}{\pi}} \left | j \right \rangle \left | k \right \rangle
\end{equation}

\subsection{The SWAP gate}
Before getting to the QFT we need one more piece of the puzzle, the \textbf{\textit{SWAP}}-gate. The gate swaps the states of two qudits and its effect can be written as:

\begin{equation}
  SWAP \left | j \right \rangle  \left | k \right \rangle = \left | k \right \rangle  \left | j \right \rangle
\end{equation}

Previous research already brought some insights on its properties and definition for arbitrary $d$-levels qudits \cite{wilmott2008construction, pourkia2023note}. Although there are multiple ways to implement it, for $d > 2$, all of them are slightly more complex than for $d = 2$ \cite{garcia2013swap} and make use of additional gates. The implementation proposed here makes use of the previously defined \textbf{\textit{SUMX}}-gate and an additional gate, that in \cite{garcia2013swap} has been called \textbf{X}, but here we are going to call it the \textbf{\textit{K}}-gate or \textbf{\textit{Complement}} operator. For $d = 2$, $K_2 \equiv I$, this is why it's not needed for the qubit case. Its matrix form and effect are the following:

\begin{equation}
  K_d \doteq \begin{bmatrix}
    1 & 0 & \ldots & 0 & 0 \\
    0 & 0 & \ldots & 0 & 1 \\
    0 & 0 & \ldots & 1 & 0 \\
    \vdots & \vdots & \ddots & \vdots & \vdots \\
    0 & 1 & \ldots & 0 & 0 \\
  \end{bmatrix}
\end{equation}
\begin{equation}
  K_d \left | j \right \rangle \doteq \left | - j \right \rangle
\end{equation}

\begin{figure}[!h]
  \centering
  \begin{adjustbox}{width=0.5\textwidth}
    \begin{quantikz}
      \lstick{$q_0$} & \gate{SUMX^{\dagger}} & \ctrl{1} & \gate{SUMX^{\dagger}} & \gate{K} &\\
      \lstick{$q_1$} & \ctrl{-1} & \gate{SUMX} & \ctrl{-1} & &
    \end{quantikz}
  \end{adjustbox}
  \caption{Schema of a qudit Swap.}
  \label{swap-schema}
\end{figure}
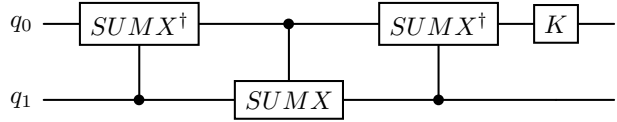

The practical effect is that, given a qudit state, it returns the complemental one$\mod d$. Given this, we can define the \textbf{\textit{SWAP}}-gate with the circuit in Fig. \ref{swap-schema}.

\section{QFT implementation}
With respect to the QFT definitions and implementations, there are already many resources for mathematical definitions and practical implementations for qubits circuits \cite{hales2000improved, coppersmith2002approximate, ruiz2017quantum, qiskit2024}. Nevertheless, in recent years, despite many gate definitions have been missing from the literature, numerous propositions have been made in order to extend them to the general qudit scenario \cite{wang2020qudits, pavlidis2017arithmetic, pavlidis2021quantum}. In particular, matrix definitions of the controlled rotations were absent. Our approach in this work has been to extrapolate the definition of a new gate, the \textbf{\textit{SUMP}}, in order to actually define and verify the QFT algorithm for a general $d$-levels qudit.

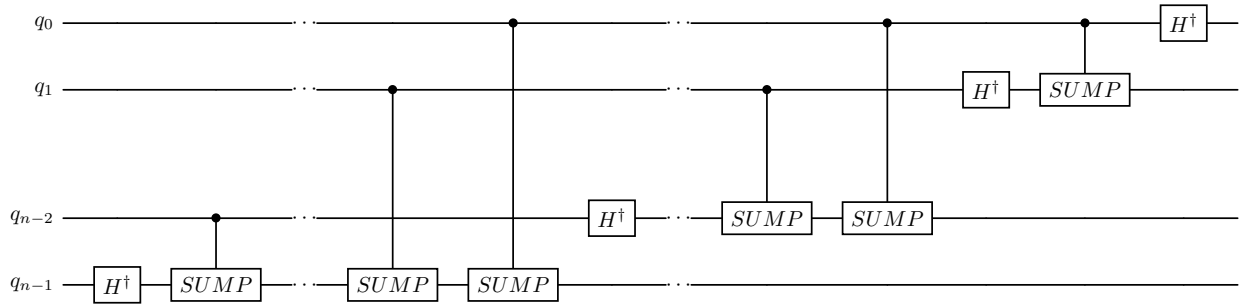
\begin{figure}
\centering
\begin{adjustbox}{width=\textwidth}
  \begin{quantikz}
    \lstick{$q_0$} & & & \cdots & & \ctrl{5} & & \cdots & & \ctrl{4} & & \ctrl{1} & \gate{H^{\dagger}} & \\
    \lstick{$q_1$} & & & \cdots & \ctrl{4} & & & \cdots & \ctrl{3} & & \gate{H^{\dagger}} & \gate{SUMP} & & \\
    \\
    \\
    \lstick{$q_{n-2}$} & & \ctrl{1} & \cdots & & & \gate{H^{\dagger}} & \cdots & \gate{SUMP} & \gate{SUMP} & & & & \\
    \lstick{$q_{n-1}$} & \gate{H^{\dagger}} & \gate{SUMP} & \cdots & \gate{SUMP} & \gate{SUMP} & & \cdots & & & & & &
  \end{quantikz}
\end{adjustbox}
\caption{Schema of a qudit QFT.}
\label{qft-schema}
\end{figure}

Given all of the above, we have every piece in order to run a QFT by applying the algorithm shown in \cite{wang2020qudits}, with the \textbf{\textit{R}} blocks replaced by our \textbf{\textit{SUMP}} and by sticking with Qiskit's little endian notation, thus inverting the \textbf{\textit{H}}-gates as well. The QFT algorithm in Fig. \ref{qft-schema} shows the position of the Hadamard and controlled rotations gates, whose angles are given by equation \ref{qft-angle}\footnote{There $q_{c,t}$ means, respectively, the index of the controlled or target qudit.}. The \textbf{\textit{SWAP}}-gates are not represented as the same as their qubit counterparts \cite{qiskit2024}.

\begin{equation}
  \theta = \pi 2^{d \, \left ( q_c - q_t \right )}
  \label{qft-angle}
\end{equation}

We simulated the circuit in Qiskit by exploiting the fact that $d$ dimensional qudits can be perfectly simulated by $n$ qubits where $d = 2^n$. We found out that, with the matrix definitions presented in this paper, the overall circuit operator of Fig. \ref{qft-schema} perfectly match SciPy's $DFT^\dagger$\footnote{Simulations with $n > 4$ couldn't be carried out because of RAM constraints.}.

\section{Conclusion}
In this paper, we successfully extended quantum gate definitions to accommodate qudits of any dimension and applied them to implement the Quantum Fourier Transform (QFT). Our novel gate proposals, including the \textbf{\textit{SUMP}}-gate, fill gaps in the literature and provide a complete set of tools for qudit quantum computing by providing the needed theoretical definitions and demonstrations. This work paves the way for future research into qudit-based algorithms, error correction, and scalable quantum computing architectures.

\section{Acknowledgements}
We would like to acknowledge M. Grossi for grateful discussion and review.

\printbibliography

\end{document}